\documentstyle[12pt]{article}
\begin{document}
\title{Pulsar Braking Index and Mass Accretion}
\author{P.D. Morley \\XonTech, Inc. \\6862 Hayvenhurst Ave.
\\ Van Nuys, CA 91406}
\maketitle
\begin{abstract}
I show that the braking index, $N$, a fundamental pulsar experimental
quantity, naturally differs from the canonical value of 3 by terms which
involve mass accretion. Using the measured values of $N$ for PSR1509-58 and
PSR0531+21, I determine that for constant density neutron stars their
present mass accretion rates are $(3.10\pm.51)\times10^{-5}$ M year$^{-1}$ and
$(9.946\pm.089)\times10^{-5}$ M year$^{-1}$ respectively, where M is the
mass of each pulsar. Finally, I demonstrate that mass accretion
removes the outstanding problem of the origin of the big glitches of the
Vela Pulsar.
\end{abstract}

\vspace{.25in}

I show that the braking index, $N$, a fundamental pulsar experimental
quantity, naturally differs from the canonical value of 3 by terms which
involve mass accretion. Using the measured values of $N$ for PSR1509-58 and
PSR0531+21, I determine that for constant density neutron stars their
present mass accretion rates are $(3.10\pm.51)\times10^{-5}$ M year$^{-1}$ and
$(9.946\pm.089)\times10^{-5}$ M year$^{-1}$ respectively, where M is the
mass of each pulsar. Finally, I demonstrate that mass accretion
removes the outstanding problem of the origin of the big glitches of the
Vela Pulsar.
\end{abstract}

\vspace{.25in}

I show that the braking index, $N$, a fundamental pulsar experimental
quantity, naturally differs from the canonical value of 3 by terms which
involve mass accretion. Using the measured values of $N$ for PSR1509-58 and
PSR0531+21, I determine that for constant density neutron stars their
present mass accretion rates are $(3.10\pm.51)\times10^{-5}$ M year$^{-1}$ and
$(9.946\pm.089)\times10^{-5}$ M year$^{-1}$ respectively, where M is the
mass of each pulsar. Finally, I demonstrate that mass accretion
removes the outstanding problem of the origin of the big glitches of the
Vela Pulsar.
\end{abstract}

\vspace{.25in}

I show that the braking index, $N$, a fundamental pulsar experimental
quantity, naturally differs from the canonical value of 3 by terms which
involve mass accretion. Using the measured values of $N$ for PSR1509-58 and
PSR0531+21, I determine that for constant density neutron stars their
present mass accretion rates are $(3.10\pm.51)\times10^{-5}$ M year$^{-1}$ and
$(9.946\pm.089)\times10^{-5}$ M year$^{-1}$ respectively, where M is the
mass of each pulsar. Finally, I demonstrate that mass accretion
removes the outstanding problem of the origin of the big glitches of the
Vela Pulsar.
\end{abstract}

\vspace{.25in}

\begin{center} PACS Numbers: 97.60Gb, 97.60Jd \end{center}

\vspace{.25in}

The braking index of a pulsar, $N$, is defined to be
\begin{equation} N = \nu \ddot{\nu} / \dot{\nu}^{2}  \end{equation}
where $\nu$ is the pulsar rotational frequency and dots designate
time-differentiation. To date, $N$ has been measured for only two
pulsars: PSR1509-58 where [1] $N=2.83 \pm 0.03$ and for PSR0531+21
with [2] $N=2.515 \pm 0.005$. Slowing down (braking) of a pulsar's
rotation by radiation reaction from a dipole magnetic field [3]
and from charged particle ejection [4] both give $N=3$. A possible
explanation of why $N<3$ is the radial deformation of the magnetic field
lines by an outgoing plasma [5]. However this requires that the
magnetosphere is plasma dominated, whereas it may be that vacuum conditions
actually apply. Another problem with this explanation is that it does
not naturally explain why $N$ is so close to 3 since in the limiting case
of a stellar wind, it would predict [6] $N=1$. In this paper, I advance the
idea that pulsar mass accretion [7] is the dominate mechanism changing
$N=3$ to $N<3$. I will show that this assumption allows us to determine
the mass accretion rate from the experimental value of $N$. Furthermore,
the existence of pulsar mass accretion removes the outstanding problem of
the origin of the huge glitches of the Vela Pulsar.

The torque equation governing the change of the angular frequency $\omega$
of the pulsar is
\begin{equation} \frac{d(I \omega)}{dt} = -H \omega^{3} \end{equation}
where $H$ is a positive constant and $I$ is the moment of inertia. The
exponent 3 on the RHS comes from the emission of dipole radiation and
charged particles. We now assume that $\dot{I} \neq 0 $. In
contradistinction to x-ray binaries, the solitary neutron star
should not experience additional torque due to mass accretion from the
immediate interstellar vicinity. From eq(1)
\begin{eqnarray}
N = 3 + \delta   \\
\delta =(\dot{I} \omega / I \dot{\omega}) -(\omega^{2} \ddot{I} / I
\dot{\omega}^{2})
\end{eqnarray}
and we see that $\delta < 0$. For purposes of evaluating $\delta$, we now
take the rather good approximation that the neutron star is uniform in density
which gives $I = constant \cdot M^{5/3}$. Experimental determination of $N$
involves observations over several years which is short by astronomical
evolutionary time scales. This allows us to put the present epoch mass
accretion rate $ \dot{M} = \beta =$ constant, where $M$ is the mass. We
thus derive the main equation
\begin{equation} \delta = \frac{5}{3} (\frac{\beta}{M})
\frac{\omega}{\dot{\omega}} - \frac{10}{9} (\frac{\beta}{M})^{2}
\frac{\omega^{2}}{\dot{\omega}^{2}}    \end{equation}
In Table 1, we list the properties of the two neutron stars whose braking
indexes
have been measured and the predictions of $\beta / M$ using eq(5).
It's important to point out that eq(5) does not fix the sign of $\beta $.
I argue that the immense gravitational field of the neutron star prevents
significant mass loss ($\beta <0$). Futhermore, as will
be shortly seen, only mass accretion ($\beta >0$) can explain the Vela
Pulsar glitches.

The most interesting question is the influence of mass accretion on the
origin of pulsar glitches. Reference 8 showed that pulsar glitches
satisfy a scaling law, and surprisingly, this scaling law is the same
as the Gutenberg-Richter law of earthquakes [9]. This equality of
scaling laws strongly implies that pulsar glitches are starquakes [10].
However, the starquake model is known to fail for the Vela Pulsar
whose actual time interval of glitches is the order of 10 years compared
to $>10^{5}$ years starquake prediction [11]. We want to add mass accretion to
the standard stress analysis done in the textbooks [11] to see if the
starquake model can now adaquetly describe the big Vela glitches.

The starquake model postulates that the pulsar has some initial deformation
$\epsilon_{0}$, in which state it is strain-free. As $\omega$ decreases, the
deformation $\epsilon$ increases beyond $\epsilon_{0}$ straining the crust. The
equilibrium distortion satisfies [11]
\begin{equation} \epsilon = \frac{I \omega^{2}}{4(A+B)} + \frac{B \epsilon_{0}}
{A+B}  \end{equation}
where A and B respectively measure the gravitational ($A \sim M^{5/3}$) and
elastic ($B \sim M$) energy stored in the star. Differentiation produces
\begin{equation} \dot{\epsilon}= \frac{T_{K}}{(A+B)}
\frac{\dot{\omega}}{\omega}
+ \frac{T_{K}B}{3(A+B)^{2}} \frac{\beta}{M} - \frac{2AB \epsilon_{0}}
{3(A+B)^{2}} \frac{\beta}{M}.  \end{equation}
$T_{K}$ is the rotational energy. In eq(7) the last two terms are new arising
from
$\dot{I} \neq 0$. The time $t_{sq}$ between starquakes is approximately [11]
\begin{equation} t_{sq} \approx \frac{\Delta \sigma}{\dot{\sigma}} =
- \frac{A}{B} \frac{\Delta \epsilon }{\dot{\epsilon}}  \end{equation}
where $\sigma$ is the mean stress. For the Vela Pulsar, $\dot{\omega} /
\omega = -1.39 \times 10^{-12} s^{-1}, \Delta \epsilon \approx 10^{-6}$
with the requirement $t_{sq} \sim 10$ years. Without $\dot{I} \neq 0$, it
is impossible to find an $A$, $B$ for the first term in eq(7) to satisfy
the Vela time constraint. However, with the new terms, there appears to
be an unique solution: perusal of detailed neutron star models [12]
shows that if $A/B \sim 1$ and $\epsilon_{0} \sim 10^{-2}-10^{-3}$, $t_{sq}$
is in the required range. As an example, a typical solid core model of
reference 12 has $A=15.0 \times 10^{52}$ ergs, $B=13.2 \times 10^{52}$ ergs,
$T_{K}=1.75 \times 10^{47}$ ergs; the first two terms in eq(7) are negligible
and the last term gives  $\dot{\epsilon} \approx -.17 \epsilon_{0} \beta /M$
(we see here that $\beta$ must be positive) and $t_{sq} \approx
6.8 \times 10^{-6} / (\epsilon_{0} \beta /M)$. For$^{12}$ $\epsilon_{0}
\sim 10^{-2}$, $\beta / M \sim 10^{-4} year^{-1}$
\begin{equation} t_{sq} \approx 7 years. \end{equation}

These requirements imply that the Vela Pulsar is a very large
neutron star of mass $M \simeq 1.9-2.0 M_{\odot}$ that has a solid core or
stratified solid core and is accreting matter at a high rate, comparable
or greater than the Crab Pulsar.

In conclusion, the assumption of pulsar mass accretion simultaneously can
explain experimental values of the braking index and the origin of the
giant Vela glitches. Derived present epoch accretion rates are reasonable.

\newpage
\begin{table}
\begin{tabular}{cccc}
Pulsar & N & $\omega / \dot{\omega}$ (s) & $\beta / M$ (year$^{-1}$) \\
\mbox{} & \mbox{} & \mbox{} & \mbox{} \\
PSR1509-58 & 2.83$\pm0.03$ & -9.754046 [10] & (3.10$\pm.51$) [-5] \\
PSR0531+21 & 2.515$\pm0.005$ & -7.91089 [10] & (9.946$\pm.089$) [-5]
\end{tabular}

\caption{The numbers in square parenthesis are base 10 exponents.}
\end{table}


\begin{thebibliography}{99}
\bibitem{} In this paper, pulsar mass accretion means the solitary
neutron star is accreting matter from its nearby interstellar environment.
\end{thebibliography}
\end{document}